\documentclass{mem}
\usepackage{natbib}\usepackage{txfonts}\usepackage{balance}
\usepackage{graphicx}
\usepackage[a4paper]{hyperref}
\idline{00}{000}
\begin{document}
\def\teff{$T\rm_{eff }$}
\def\kms{$\mathrm {km s}^{-1}$}

\title{
e-VLBI observations of SN2001em -- an off-axis GRB candidate
}

   \subtitle{}

\author{
Zsolt Paragi\inst{1}, Michael A. Garrett\inst{1}, Bohdan Paczy\'nski\inst{2}, 
Chryssa Kouveliotou\inst{3}, Arpad~Szomoru\inst{1}, Cormac Reynolds\inst{1}, 
Stephen M. Parsley\inst{1} \and Tapasi Ghosh\inst{4} 
} 

\offprints{Z. Paragi}

\institute{
Joint Institute for VLBI in Europe, 
Postbus 2, 7990AA Dwingeloo, 
Netherlands\newline
\email{zparagi@jive.nl}
\and
Department of Astrophysical Sciences, Princeton University, Peyton Hall, Ivy Lane, Princeton NJ 08544-1001
\and
National Space Science Technology Center, NASA/MSFC, XD-12, Huntsville, AL 35805, USA
\and
Arecibo Observatory,
HC03 Box 53995, Arecibo,
Puerto Rico 00612
}

\authorrunning{Z. Paragi}

\titlerunning{e-VLBI observations of SN2001em}

\abstract{

Studying transient phenomena with the Very Long Baseline Interferometry (VLBI) 
technique faces severe difficulties because the turnaround time of the 
experiments from the observations to the scientific result is rather long.
The e-VLBI technique has made it possible to transfer the data from a number
of European VLBI Network (EVN) telescopes to the central data processor at 
JIVE through optical fibres, and correlate them in real time. 
The main goal of this paper
is to introduce this rapidly developing new technique, by presenting observational 
results from a recent experiment. We observed SN2001em, 
a Type Ib/c supernova with an e-VLBI array and the Multi-Element Radio Linked 
Interferometer Network (MERLIN) in the UK. The source is marginally detected
in our observations. We cannot make definite conclusions whether it is 
resolved at 1.6~GHz or not. Our data show that SN2001em either started fading in
the last couple of months, or its radio spectrum is inverted at low frequencies,
indicating free-free or synchrotron self-absorption. This is quite unusual, but
not unprecedented in radio SNe. 

\keywords{Techniques: interferometry -- Stars: supernovae: individual (SN2001em) 
-- Radio continuum: general -- Gamma rays: bursts} 
}
\maketitle{}

\section{Introduction}

There have been significant developments in the EVN in the recent years. Among these,
the most substantial is the replacement of the MarkIV tape-based recording system
with the Mark\,5 \cite{mk5, whitney} disk-based recording system. The advantages 
of disk recording are improved recording performance, and the possibility of monitoring 
the stations during the observations (e.g. transfer of small amounts of telescope data via the
Internet, the so-called ftp fringe tests), which allows more robust
operations. There is ongoing development to make the EVN operationally capable of 
electronic VLBI (e-VLBI), where the data are transfered to the correlator through 
optical fibres, and correlated in real time \cite{mag1, mag2, steve, arpad}, 
without any intermediate recording on disk. 


e-VLBI -- especially when combined with the 
data calibration pipeline \cite{rey02} -- offers a unique 
opportunity to carry out target-of-opportunity style experiments,
with a very quick science turnaround time that has been hitherto unavailable in 
VLBI. We demonstrate this by presenting e-VLBI observations of 
the faint supernova SN2001em with the EVN. A full discussion of these observations
will be presented in Garrett et al. (2005, in prep.). 

\section{SN2001em}

On 15 September 2001, Papenkova \cite{discovery} discovered SN2001em in UGC\,11794,
a nearby galaxy at a distance of 80~Mpc. The early spectrum indicated a Type Ib/c
SN \cite{type}, but later broad H$\alpha$ lines appeared \cite{broad} that are 
not typical of this class.
Radio and X-ray emission from the source was detected only two years after the
initial explosion. The radio flux density increased from 1.15 to 1.48 mJy between
October 2003 and January 2004, as measured by the VLA \cite{vla}.
The spectral index was steep, $\alpha=-0.36\pm0.16$ ($S\propto\nu^{\alpha}$), 
indicating optically thin synchrotron emission. 
In the X-rays the source was detected with Chandra with a 0.5-8 keV luminosity of 
$\sim10^{41}$erg\,s$^{-1}$ \cite{chandra}.

It is puzzling why the radio and X-ray emission came so late, especially at the 
observed high luminosities. Granot and Ramirez-Ruiz \cite{jet} investigated two
scenarios: {\it a)} the emission comes from interaction of the SN shell with the 
circumstellar medium, or, {\it b)} it originates in a mildly relativistic jet. In the latter
case the jet Lorentz factor is initially very high ($\Gamma_{0}\geq2$), and 
the emission is highly beamed. For an observing line of sight
far from the direction of the jet, the emission is strongly de-boosted.
As a result, the jet is initially not observed, and only 
becomes apparent when it decelerates to mildly
relativistic speeds. Granot and Ramirez-Ruiz \cite{jet} ruled out the shell-interaction
scenario in favor of the jet model. 

The idea that Type Ib/c SNe may be related to relativistic jet GRBs originally came
from Paczy\'nski \cite{paci}. Long duration GRBs tend to appear near star-forming 
regions,  which indicates a supernova origin. Paczy\'nski \cite{paci} estimated that 
one out of 100-1000 core-collapse
supernovae generate ultra-relativistic jets, but only the ones that are aligned with
our line of sight produce GRBs. Nearby SNe that show late time radio emission (possibly
related to a decelerated jet) could be resolved with the VLBI technique \cite{grbjet}.

\section{Observations}

\begin{table*}
\caption{Parameters of the participating telescopes: aperture, diameter, system 
temperature, and the maximum capacity of the Internet or dedicated fibre connection. 
Baseline sensitivities to Arecibo are also given, 
assuming 10 minutes integration time, and 64 Mbps data rate. 
At the time of the observations,
Arecibo was limited to 100~Mbps. The Cambridge data rate was
limited by the microwave link between the station and Jodrell Bank.}
\label{array}
\begin{center}
\begin{tabular}{lrrrr}
\hline
\\
Telescope    & Diameter & $T_{\rm sys}$ & Maximum capacity  & Baseline sensitivity \\
\hline
\\
Arecibo      &    305m  &       3K      &          100 Mbps &         $-$          \\
Cambridge    &     32m  &     212K      &    $\sim$112 Mbps &      364$\mu$Jy      \\
Jodrell Bank &     76m  &      44K      &         1024 Mbps &      165$\mu$Jy      \\
Onsala       &     25m  &     390K      &         1024 Mbps &      493$\mu$Jy      \\
Torun        &     32m  &     230K      &         1024 Mbps &      379$\mu$Jy      \\
Westerbork   &  14x25m  &      30K      &         1024 Mbps &      136$\mu$Jy      
\\
\hline
\end{tabular}
\end{center}
\end{table*}


We observed SN2001em on 11 March 2005 at a frequency of 1.6\,GHz (corresponding to 18\,cm 
in wavelength), with Arecibo, Cambridge, Jodrell Bank, Onsala, Torun
and Westerbork (see Table~\ref{array}). The data were transfered to the EVN Data Processor at JIVE 
through optical fibres, using the pan-European research network G\'EANT (http://www.geant.net/),
and correlated in real time. Initially we observed fringe-finder sources and corrected for clock
errors. At this early phase we observed at 128 Mbps data rate with the European telescopes only. 
The data rate was
lowered to 64 Mbps when Arecibo joined the observations. The target was phase-referenced to
J2145+1115 (1.4 degrees away), in an 11 minutes switching cycle. During the whole experiment the data 
quality was monitored and real time fringe plots were made available to the telescope operators 
through the Internet. All stations produced good data. In parallel to the e-VLBI recording, the MERLIN 
array in the UK also observed the source.

The weather conditions were not ideal during this experiment. The Lovell Telescope at Jodrell Bank
had to be stopped soon after the start because of high winds. Some hours later the Cambridge
telescope was also lost for the same reason, before the target source observations had started.
Additionally, there were other problems like severe interference at Arecibo at the end of the experiment.
Eventually we collected limited, but valuable data on SN2001em. A detailed description
of these observations, together with a thorough review of the e-VLBI developments at the EVN will be 
published by Garrett et al. (2005, in prep.).

\section{Data reduction and results}

The uv-FITS file was prepared right after the observations, and the data were immediately pipelined.
The FITS file, AIPS calibration tables, and the pipeline plots are publicly available from the
EVN Data Archive (http://archive.jive.nl/, experiment IG002B). The target source was
not visible on the dirty map produced by the EVN pipeline. But since the MERLIN data showed the
target as a $\sim$880~$\mu$Jy unresolved source, we took a closer look at the e-VLBI data in Difmap.

In phase-referencing most but not all of the atmospheric phase errors are corrected by using a nearby 
reference source. However, there are systematic phase errors that arise during the transfer of solutions 
from one source to the other, due to the limitations of the correlator model. This may be corrected for 
by self-calibrating the target, with solution intervals much longer than the initial coherence time.
Indeed, we detected SN2001em after self-calibrating with a 30 minutes solution interval (assuming a
point source model). This was further improved by correcting for Arecibo phases using 10 minutes 
self-calibration intervals. The phases at other stations were kept fixed, because only
the Westerbork-Arecibo baseline had sufficient sensitivity for self-calibration with shorter
solution intervals. 

\begin {figure}[]
\resizebox{\hsize}{!}{\includegraphics[bb=33 140 582 721, clip=true]{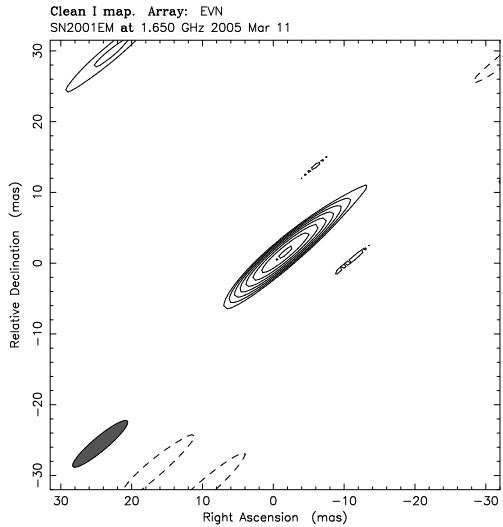}}
\caption{\footnotesize
Naturally weighted map of SN2001em in our e-VLBI experiment. The peak brightness is 
529~$\mu$Jy/beam using a restoring beam of 15x2 mas (slightly over-resolved), with 
major axis PA at $-$50 degrees. The contour levels are -1, 1, $\sqrt(2)$, 2 etc. times
44~$\mu$Jy/beam.
}
\label{sn2001em}
\end{figure}

SN2001em is marginally detected with a signal-to-noise ratio of 4--5 (after the 30~minutes self-calibration), 
with a correlated flux density of 571~$\mu$Jy/beam on trans-Atlantic baselines (Fig.~\ref{sn2001em}). 
This is significantly fainter than measured by MERLIN, even if one assumes a conservative 10\% accuracy
level for amplitude calibration of both arrays. One may argue that phase-referencing failed, and the
10 minute solution time in self-calibration was not short enough to recover the Arecibo phases properly.
However we carried out direct fringe-fitting on the Westerbork-Arecibo baseline and got only occasional,
very weak solutions, which is unexpected for a $~$1\,mJy compact source (unless Arecibo and/or Westerbork 
was much less sensitive than expected -- the calibrator data show that this is not the case). This 
indicates that SN2001em
is slightly resolved in our observations. We performed model fitting to the uv~data. Because of the very 
limited dataset, we constrained the Gaussian model to be circular. Our result is 
a 900~$\mu$Jy component (in perfect agreement with the MERLIN results) with a size of 2.45~mas. A unifomly
bright disk model was also fitted, which resulted in a size of 1.09~mas. Due to the limited amount of data
and the low-SNR detection, the error in our size measurement is rather large (about the size of the beam, 
with a 2.44 milliarcsecond minor axis). 

\section{Discussion}

SN2001em had been observed by two other independent groups before our experiment. Both projects were
carried out at 8.4~GHz, a significantly higher frequency than used in our observations. 
Stockdale et al. \cite{vlba} measured $1.8\pm0.2$~mJy flux density with the VLBA on 1 July 2004, and 
the source appeared unresolved with a resolution of $1.9\times0.8$~mas. Bietenholz and Bartel \cite{hsa} 
used the High Sensitivity Array (a global
array consisting of the VLBA, Arecibo, GBT, phased VLA and Effelsberg) on 22 November 2004. They detected a
$1.5\pm0.1$~mJy, practically unresolved source, with a $3\sigma$ upper limit of $0.35-0.59$~mas, depending on
the type of model fitted. This indicates that the size measured in our experiment indeed has a large error,
unless the lower frequency flux originates in a different emitting region.

It should be noted that the flux measured in our experiment is significantly lower than the earlier
measurements indicate, which showed that the source had been permanently brighter than 1~mJy since its 
detection. Either SN2001em has faded in the last couple of months, or the spectrum between 1.6-8.4~GHz is 
inverted. The spectral index reported earlier was $\alpha=-0.36\pm0.16$ \cite{vla} between 5 and 15~GHz. 
If the flux was constant, our data would indicate a spectral index of $\alpha\sim+0.3$. Although this is 
atypical of radio SNe, it is not unprecedented. Bietenholz et al. \cite{sn1986j} detected an inverted 
spectrum compact component in SN1986J, 20 years after the explosion. It may be that these inverted spectrum 
compact sources in SNe are related to a newly born plerion nebula around a young pulsar, but the jet origin 
cannot be fully excluded yet.

\section{Conclusions}

We detect SN2001em, a sub-mJy source that is two orders of magnitude fainter than the other objects 
observed by e-VLBI to date. Our attempts at measuring the source size remain inconclusive, but we 
show that SN2001em either faded, or has a spectrum that is inverted at lower frequencies, possibly 
indicating free-free or synchrotron self-absorption. Such an inverted spectrum radio component in 
SNe is not unprecedented. 

The main results of this experiment became available within a couple of days after the observations. 
With the aid of e-VLBI, high-resolution radio images of e.g. transient sources can be made available in 
a very short turnaround time. We believe that this will change the impact VLBI has in contributing to the 
study of transient phenomena.

\begin{acknowledgements}
We are grateful to Peter Thomasson for quickly pipelining the MERLIN
data for us.
The European VLBI Network is a joint facility of European, Chinese,
South African and other radio astronomy institutes funded by their
national research councils. The Arecibo Observatory is part of the 
National Astronomy and Ionosphere Center, which is operated by 
Cornell University under a cooperative agreement with National Science
Foundation, NSF.
\end{acknowledgements}

\bibliographystyle{aa}

\end{document}